\documentclass[twoside,twocolumn,9pt]{article}
\usepackage{extsizes}
\usepackage[super,sort&compress,comma]{natbib} 
\usepackage[version=3]{mhchem}
\usepackage[left=1.5cm, right=1.5cm, top=1.785cm, bottom=2.0cm]{geometry}
\usepackage{balance}
\usepackage{times,mathptmx}
\usepackage{sectsty}
\usepackage{graphicx} 
\usepackage{lastpage}
\usepackage[format=plain,justification=justified,singlelinecheck=false,font={stretch=1.125,small,sf},labelfont=bf,labelsep=space]{caption}
\usepackage{float}
\usepackage{fancyhdr}
\usepackage{fnpos}
\usepackage[english]{babel}
\usepackage{array}
\usepackage{droidsans}
\usepackage{charter}
\usepackage[T1]{fontenc}
\usepackage[usenames,dvipsnames]{xcolor}
\usepackage{setspace}
\usepackage[compact]{titlesec}

\usepackage{epstopdf}

\definecolor{cream}{RGB}{222,217,201}

\begin{document}

\pagestyle{fancy}
\thispagestyle{plain}
\fancypagestyle{plain}{

\fancyhead[C]{\includegraphics[width=18.5cm]{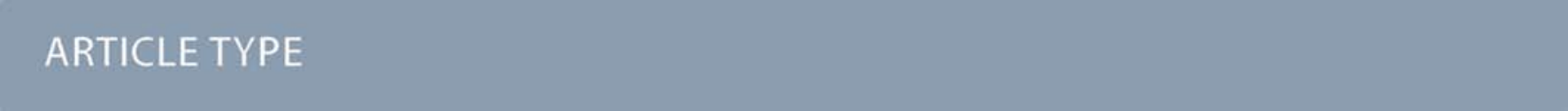}}
\fancyhead[L]{\hspace{0cm}\vspace{1.5cm}\includegraphics[height=30pt]{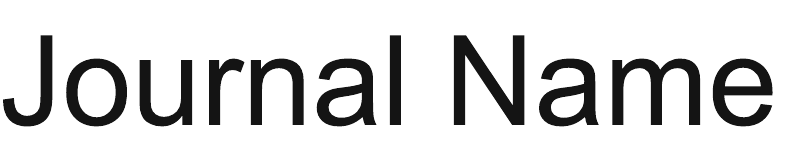}}
\fancyhead[R]{\hspace{0cm}\vspace{1.7cm}\includegraphics[height=55pt]{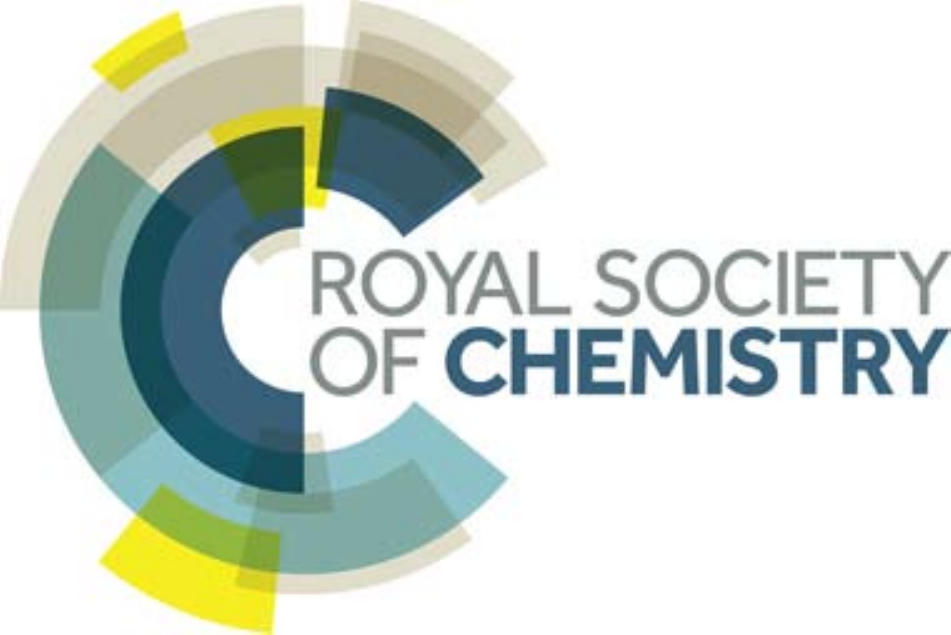}}
\renewcommand{\headrulewidth}{0pt}
}

\makeFNbottom
\makeatletter
\renewcommand\LARGE{\@setfontsize\LARGE{15pt}{17}}
\renewcommand\Large{\@setfontsize\Large{12pt}{14}}
\renewcommand\large{\@setfontsize\large{10pt}{12}}
\renewcommand\footnotesize{\@setfontsize\footnotesize{7pt}{10}}
\makeatother

\renewcommand{\thefootnote}{\fnsymbol{footnote}}
\renewcommand\footnoterule{\vspace*{1pt}%
\color{cream}\hrule width 3.5in height 0.4pt \color{black}\vspace*{5pt}} 
\setcounter{secnumdepth}{5}

\makeatletter 
\renewcommand\@biblabel[1]{#1}            
\renewcommand\@makefntext[1]%
{\noindent\makebox[0pt][r]{\@thefnmark\,}#1}
\makeatother 
\renewcommand{\figurename}{\small{Fig.}~}
\sectionfont{\sffamily\Large}
\subsectionfont{\normalsize}
\subsubsectionfont{\bf}
\setstretch{1.125} 
\setlength{\skip\footins}{0.8cm}
\setlength{\footnotesep}{0.25cm}
\setlength{\jot}{10pt}
\titlespacing*{\section}{0pt}{4pt}{4pt}
\titlespacing*{\subsection}{0pt}{15pt}{1pt}

\fancyfoot{}
\fancyfoot[LO,RE]{\vspace{-7.1pt}\includegraphics[height=9pt]{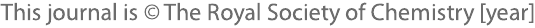}}
\fancyfoot[CO]{\vspace{-7.1pt}\hspace{13.2cm}\includegraphics{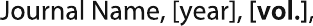}}
\fancyfoot[CE]{\vspace{-7.2pt}\hspace{-14.2cm}\includegraphics{RF}}
\fancyfoot[RO]{\footnotesize{\sffamily{1--\pageref{LastPage} ~\textbar  \hspace{2pt}\thepage}}}
\fancyfoot[LE]{\footnotesize{\sffamily{\thepage~\textbar\hspace{3.45cm} 1--\pageref{LastPage}}}}
\fancyhead{}
\renewcommand{\headrulewidth}{0pt} 
\renewcommand{\footrulewidth}{0pt}
\setlength{\arrayrulewidth}{1pt}
\setlength{\columnsep}{6.5mm}
\setlength\bibsep{1pt}

\makeatletter 
\newlength{\figrulesep} 
\setlength{\figrulesep}{0.5\textfloatsep} 

\newcommand{\topfigrule}{\vspace*{-1pt}%
\noindent{\color{cream}\rule[-\figrulesep]{\columnwidth}{1.5pt}} }

\newcommand{\botfigrule}{\vspace*{-2pt}%
\noindent{\color{cream}\rule[\figrulesep]{\columnwidth}{1.5pt}} }

\newcommand{\dblfigrule}{\vspace*{-1pt}%
\noindent{\color{cream}\rule[-\figrulesep]{\textwidth}{1.5pt}} }

\makeatother

\twocolumn[
  \begin{@twocolumnfalse}
\vspace{3cm}
\sffamily
\begin{tabular}{m{4.5cm} p{13.5cm} }

\includegraphics{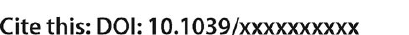} & \noindent\LARGE{Quantum tunneling during interstellar surface-catalyzed formation of water: the reaction \ce{H + H2O2 -> H2O + OH}} \\
\vspace{0.3cm} & \vspace{0.3cm} \\

 & \noindent\large{Thanja Lamberts,$^{\ast}$\textit{$^{a}$} Pradipta Kumar Samanta,\textit{$^{a}$} Andreas K\"ohn,\textit{$^{a}$}  and Johannes K\"astner\textit{$^{a}$}} \\

\includegraphics{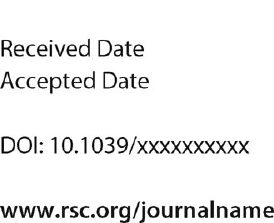} & \noindent\normalsize{The final step of the water formation network on interstellar grain surfaces starting from the \ce{H + O2} route is the reaction between H and \ce{H2O2}. This reaction is known to have a high activation energy and therefore at low temperatures it can only proceed via tunneling. To date, however, no rate constants are available at temperatures below 200~K. 
In this work, we use instanton theory to compute rate constants for the title reaction with and without isotopic substitutions down to temperatures of 50~K.
The calculations are based on density functional theory, with additional benchmarks for the activation energy using unrestricted single-reference and multireference coupled-cluster single-point energies. Gas-phase bimolecular rate constants are calculated and compared with available experimental data not only for \ce{H + H2O2 -> H2O + OH}, but also for \ce{H + H2O2 -> H2 + HO2}. We find a branching ratio where the title reaction is favored by at least two orders of magnitude at 114~K. In the interstellar medium this reaction predominantly occurs on water surfaces, which increases the probability that the two reactants meet. To mimic this one, two, or three spectator \ce{H2O} molecules are added to the system. Eley-Rideal bimolecular and Langmuir-Hinshelwood unimolecular rate constants are presented here. The kinetic isotope effects for the various cases are compared to experimental data as well as to expressions commonly used in astrochemical models. Both the rectangular barrier and the Eckart approximations lead to errors of about an order of magnitude. Finally, fits of the rate constants are provided as input for astrochemical models. 
} \\

\end{tabular}

 \end{@twocolumnfalse} \vspace{0.6cm}

    ]


\renewcommand*\rmdefault{bch}\normalfont\upshape
\rmfamily
\section*{}
\vspace{-1cm}


\footnotetext{$^\ast$ Corresponding author}
\footnotetext{\textit{$^{a}$~Institute for Theoretical Chemistry, University of Stuttgart, Stuttgart, Germany. Tel: +49 (0)711 685 64833; E-mail: lamberts@theochem.uni-stuttgart.de}}





\section{Introduction}
In the dense and cold regions of the Interstellar Medium (ISM), water is known to be formed on the surface of dust grains via sequential hydrogenation of O, \ce{O2}, or \ce{O3}. The full water surface reaction network consists of $\sim15$ reactions and depending on density, temperature and H, \ce{H2}, and O abundance of the interstellar region, different reaction pathways towards the formation of water are important.\citep{Cuppen:2007A,Lamberts:2014} In regions with the highest density, the absolute amount of oxygen becomes sufficiently high for the following reaction pathway to contribute significantly to the formation of \ce{H2O} and OH: 
\[  \text{O} \xrightarrow{\text{O}} \text{O}_2 \xrightarrow{\text{H}} \text{HO}_2 \xrightarrow{\text{H}} \text{H}_2\text{O}_2 \xrightarrow{\text{H}} \text{H}_2\text{O} + \text{OH} .\]
The final step in this reaction route is known to have a large activation barrier in the gas phase, see \citet{Baulch:2005} and references therein. It has also been studied experimentally in the solid phase at low temperatures and a kinetic isotope effect was found.\citep{Oba:2014} This indicates the importance of tunneling which allows it to be efficient even at 15~K. Note that the surface reaction takes place in an environment that consists predominantly of water molecules. Since the \ce{H2O2} molecule can partake in several hydrogen bonds, it is expected that there is an influence of the surface on the course of the reaction. Another important influence of the surface is the increased concentration of reactants as well as heat dissipation of the exothermicity of the reaction. 
Furthermore, the gas-phase detections and non-detections of \ce{H2O2} in a diverse sample of sources\citep{Bergman:2011,Parise:2014,Liseau:2015} gave rise to the conclusion that the production of peroxide and therefore the gas-phase detectability is very sensitive to temperature. The surface destruction of peroxide was taken into account by rescaling the reaction rate according to experimental data.\citep{Du:2012,Parise:2014} Quantitatively rescaling a rate is, however, not trivial for a reaction that is deeply embedded within a reaction network.

There are two reactions possible between H and \ce{H2O2}:the first, ~\ref{R1}, where the O--O bond is broken and a second,~\ref{R2}, where a H-atom is abstracted.
\begin{align}
\ce{H + H2O2} &\xrightarrow{k_1} \ce{H2O + OH} \;\; \tag{R1}\label{R1} \\
\ce{H + H2O2} &\xrightarrow{k_2} \ce{HO2 + H2} \;\; \tag{R2}\label{R2} \;.
\end{align}
Only the first reaction produces \ce{H2O}, but the total reaction rate constant, $k_\text{tot.}=k_1+k_2$, determines the destruction of \ce{H2O2} on the surface. Therefore, both reactions need to be considered for an accurate description of the surface process. Furthermore, since tunneling is involved, the kinetic isotope effect should also be studied explicitly. 

Here, we present a theoretical study of the reactions~\ref{R1} and~\ref{R2} within the concept of studying reactions on a surface that proceed via tunneling. Calculations are performed using a DFT functional and basis set combination that is benchmarked to single-reference and multireference coupled-cluster single-point energies as outlined in Sections~\ref{DFTbenchmark} and ~\ref{RDDFT}. Rate constants are calculated with instanton theory which is briefly described in Section~\ref{RateConstants}. We give activation barriers and rate constants for three different cases: the gas-phase reaction (Section~\ref{Gas}), the reaction with several spectator \ce{H2O} molecules (small clusters) to mimic a surface related to a bimolecular reaction (Section~\ref{ER}), and the reaction with the same small clusters related to a unimolecular reaction (Section~\ref{LH}). Previous studies have focused only on the pure gas-phase reaction, did not take into account the differences between hydrogenation and deuteration, have not benchmarked their DFT functional, neglected tunneling and/or calculated rate constants only down to 200~K.\citep{Koussa:2006,Ellingson:2007,Taquet:2013} Finally we explain how the calculated reaction rate constants can be implemented in astrochemical models (Section~\ref{AstroIm}) and give more general conclusions (Section~\ref{Concl}).

\section{Methods}

\subsection{Electronic structure}\label{DFTbenchmark}
{The core system is small (19 electrons) and could be well treated by high-accuracy methods, but the actual focus is a reaction on a surface} of water molecules and hence the method of choice to describe the electronic structure is density functional theory (DFT). A suitable functional and basis set needs to describe the interaction between hydrogen and hydrogen peroxide, as well as between water molecules. We perform a benchmark study {for the activation and reaction energies of reactions~\ref{R1} and ~\ref{R2}} with respect to two coupled cluster methods. {The first benchmark method is  spin-orbital based (unrestricted) coupled-cluster theory with singles, doubles, and perturbative triples clusters (UCCSD(T)) \citep{Knowles:1993, Knowles:err, Deegan:1994} and explicitly-correlated geminal functions (UCCSD(T)-F12)\citep{Adler:2007, Knizia:2009} employing a restricted Hartree-Fock (RHF) reference function and the cc-pVTZ\citep{Dunning:1989} or cc-pVTZ-F12\citep{Peterson:2008} basis set. To test for multireference character, the internally contracted multireference coupled-cluster method, again with singles, doubles, and perturbative triples clusters (icMRCCSD(T)) was used.\citep{Hanauer:2011, Hanauer:2012} These computations were carried out with a cc-pVTZ basis set.\citep{Dunning:1989} A complete active space self-consistent field (CASSCF) reference was used for this method. The CAS used for \ce{H2O2}, {the transition state}, \ce{H2O}, \ce{OH}, and \ce{HO2} are (6e,6o), (7e, 7o), (4e,4o), (3e,3o), and (5e,5o). The active space was chosen to describe the unpaired electron and all bonding and antibonding sigma orbitals of the system. These benchmark computations were carried out for single geometries, as obtained from DFT computations optimized on MPW1B95/MG3S\citep{Zhao:2004, Lynch:2003} and M05-2X/MG3S\citep{Zhao:2006, Lynch:2003} levels for reactions~\ref{R1} and~\ref{R2}, respectively. This choice of functional {is based on finding the best match to coupled cluster energetics. Data are given in} Section~\ref{RDDFT}.}

{Additionally, the reaction energies are also compared to the high-accuracy extrapolated ab initio thermochemistry (HEAT) theoretical model that has been shown to go beyond CCSD(T) methods in accuracy.\citep{Harding:2008}}

{For the benchmark, we follow the approach of \citet{Ellingson:2007} and construct a set of commonly or previously\citep{Ellingson:2007, Taquet:2013} used functionals (BHLYP\citep{Becke:1988, Becke:1993a, Lee:1988}, B3LYP\citep{Becke:1988, Becke:1993, Lee:1988}, PBE0\citep{Perdew:1996, Adamo:1999}, PWB6K\citep{Zhao:2005}, MPW1B95\citep{Zhao:2004}, M05-2X\citep{Zhao:2006}) in combination with the basis sets def2-TZVPD\citep{Weigend:1998, Rappoport:2010} and MG3S\cite{Lynch:2003, Zheng:2011}. The MG3S basis set is equivalent to 6-311+G(3d2f,2df,2p) for H and O atoms. }

All geometry optimizations are performed using DL-find\citep{Kaestner:2009} within the Chemshell\citep{Sherwood:2003, Metz:2014} framework and NWChem versions 6.3 and 6.6\citep{Valiev:2010}. The single energy points are calculated with Molpro\citep{MOLPRO} for UCCSD(T) and UCCSD(T)-F12 and GeCCo\citep{Hanauer:2011} for icMRCCSD(T). VMD version 1.9.2\citep{VMD} and wxMacMolPlt version 7.7\citep{wxM} are used for visualization.

\subsection{Reaction rate constants}\label{RateConstants}
The hydrogenation reactions are initially modeled in the gas phase. To investigate the influence of the surface more specifically, we added one, two, or three water molecules to the previously optimized structures and re-optimized the resulting configuration. Reaction rate constants are calculated using instanton theory \citep{Langer:1967, Langer:1969, Miller:1975, Callan:1977, Coleman:1977, Gildener:1977, Affleck:1981, Coleman:1988, Hanggi:1990, Benderskii:1994, Messina:1995, Richardson:2009, Kryvohuz:2011, Althorpe:2011, Rommel:2011, Rommel:2011b, Kryvohuz:2014, Zhang:2014, Richardson:2016} which has been shown to provide accurate tunneling rates down to very low temperature and is increasingly used to predict rate constants.\cite{Chapman:1975, Mills:1994, Mills:1995, Mills:1997, Siebrand:1999, Smedarchina:2003, Qian:2007, Andersson:2009, Goumans:2010a, Andersson:2011, Goumans:2011c, Goumans:2011b, Rommel:2011, Goumans:2010, Jonsson:2010, Meisner:2011, Goumans:2011a, Einarsdottir:2012, Rommel:2012, Kryvohuz:2012, Kaestner:2013, Alvarez:2014, Kaestner:2014, Kryvohuz:2014, Meisner:2016} 

\begin{table*}[t!]
 \centering
 \caption{DFT functional/basis set combination benchmark with respect to UCCSD(T)-F12/cc-pVTZ-F12, UCCSD(T)/cc-pVTZ and frozen-core icMRCCSD(T)/cc-pVTZ single-point energies for reactions \ref{R1} and \ref{R2}, respectively. Reaction energies computed from the HEAT protocol are given, too. Values are given in kJ/mol excluding zero-point energies.}\label{benchmark}
 \begin{tabular}{llllll}
\hline
 & Method Ref. & \multicolumn{2}{l}{\ce{H + H2O2 -> H2O + OH}} & \multicolumn{2}{l}{\ce{H + H2O2 -> HO2 + H2}} \\
 & & Activation energy & Reaction energy & Activation energy & Reaction energy \\
\hline
UCCSD(T)-F12/cc-pVTZ-F12 & \citep{Adler:2007, Knizia:2009, Peterson:2008} & 25.5 & -299.3 & 39.4 & -66.6 \\
UCCSD(T)/cc-pVTZ & \citep{Knowles:1993, Knowles:err, Deegan:1994, Dunning:1989} & 27.7 & -294.3 & 39.6 & -69.8 \\
icMRCCSD(T)/cc-pVTZ & \citep{Hanauer:2011, Hanauer:2012, Dunning:1989} & 24.9 & -292.2 & 38.3 & -70.9 \\
HEAT-456QP & \citep{Harding:2008} &  & -297.7 & & -66.5 \\
\hline
BHLYP/def2-TZVPD & \citep{Becke:1988, Becke:1993a, Lee:1988, Weigend:1998, Rappoport:2010} & 27.2 & -331.4 & 27.6 & -89.2 \\
B3LYP/def2-TZVPD & \citep{Becke:1988, Becke:1993, Lee:1988, Weigend:1998, Rappoport:2010} & 10.8 & -299.3 & 7.3 & -90.3 \\
B3LYP/MG3S & \citep{Becke:1988, Becke:1993, Lee:1988, Lynch:2003} & 11.2 & -300.2 & 8.1 & -88.1 \\
PBE0/def2-TZVPD & \citep{Perdew:1996, Adamo:1999, Weigend:1998, Rappoport:2010} & 20.7 & -288.0 & 17.3 & -74.4 \\
PBE0/MG3S & \citep{Perdew:1996, Adamo:1999, Lynch:2003} & 21.4 & -289.0 & 18.1 & -72.3 \\
PWB6K/MG3S & \citep{Zhao:2005, Lynch:2003} & 36.0 & -307.5 & 35.4 & -74.2 \\
MPW1B95/MG3S & \citep{Zhao:2004, Lynch:2003} & 26.5 & -291.8 & 23.7 & -76.7 \\
M05-2X/MG3S & \citep{Zhao:2006, Lynch:2003} & 45.9 & -303.3 & 39.7 & -68.5 \\
\hline
\end{tabular}
\end{table*}

Instanton theory treats the quantum effects of atomic movements by Feynman path integrals.  {The main tunneling path, the instanton, is described by a closed Feynman path, which connects the reactant and product valleys of the potential energy surface.  The instanton represents the tunneling path with the highest statistical weight at a given temperature. It is located by a Newton--Raphson optimization scheme.\citep{Rommel:2011, Rommel:2011b} A semiclassical approximation results in the rate constants. More details on our implementation of instanton theory are given elsewhere.\citep{Rommel:2011, Rommel:2011b}
Rotational and translational partition functions were approximated by their classical analogues {(J-shifting approximation)}, which is generally accepted as a good approximation at the temperature scale considered here. In bimolecular cases, the product of the partition functions of the separated reactants was used, in unimolecular cases, the partition function of the encounter complex.} 
The Feynman paths were discretized to 60 images. Instanton theory is applicable below the crossover temperature $T_\text{c}$, which is defined as
\begin{equation} T_\text{c} = \frac{\hbar \omega_b}{2\pi k_\text{B}} \end{equation} where $\omega_b$ is the absolute value of the imaginary frequency at the transition state.  The instanton represents the tunneling path with the highest statistical weight at a given temperature. Instantons were optimized to a residual  gradient {(derivative of the effective energy of the instanton with respect to the mass-weighted atomic coordinates)} below $10^{-8}$ atomic units ({hartree bohr$^{-1}$ m$_\text{e}^{-1/2}$}). This and other parameters were chosen equivalently to previous work.\citep{Rommel:2011, Rommel:2011b}

First, we consider the pure gas-phase reaction. This may not be very relevant in terms of astrochemistry, but it is important to understand the simplest case first. In the gas phase, a hydrogen atom can approach the molecule on both oxygen atoms, since they are equivalent. In other words, the rate constants need to be multiplied with a rotational symmetry factor. The symmetry factor used here is 2, resulting from the pointgroup $C_2$ for the hydrogen peroxide molecule.\citep{Fernandez:2007}
Concerning the surface reaction, there are typically two reaction mechanisms taken into consideration: the Eley-Rideal (ER) and the Langmuir-Hinshelwood (LH) processes. The ER mechanism describes one species to be adsorbed on the surface and the other approaching from the gas phase, \emph{i.e.}, an overall bimolecular reaction. For the LH mechanism both species are adsorbed on the surface, approach each other via diffusion and form an encounter complex of H and \ce{H2O2} on the surface. This encounter, or pre-reactive, complex can then decay to yield the reaction products in a unimolecular process. Moreover, to extend the results to the solid phase it is key to realize that rotational motion on the surface is restricted. Therefore, rate constants calculated for both ER and LH mechanisms need to keep the rotational partition function constant between the reactant and transition state. Moreover, the surface structure breaks the gas-phase symmetry, hence no symmetry factor is required.

For astrochemical modelers to be able to easily implement the calculated rate constant, we fitted these to the rate expression\citep{Zheng:2010}
\begin{equation}
 k = \alpha \left(\frac{T}{300\mbox{~K}}\right)^\beta \exp\left({-\frac{\gamma (T + T_0)}{(T^2 +T_0^2)}}\right) .\label{Zheng}
\end{equation} 
The parameters $\alpha$, $\beta$, $\gamma$, and $T_0$ are all fitting parameters, where $\alpha$ has the units of the rate constant, $\beta$ regulates the low-temperature behavior, and $\gamma$ and $T_0$ can be related to the activation energy of the reaction. 
Instanton rate calculations were used for the fits at low temperature, below $T_\text{c}$, while rate constants obtained from transition state theory including quantized vibrations and a symmetric Eckart model for the barrier were used above $T_\text{c}$.

\section{Results and Discussion}
Here we first present the results of the DFT benchmark (Section~\ref{RDDFT}) and subsequently give the results for the three cases of the reaction that we study: the gas-phase reaction (Section~\ref{Gas}), the bimolecular Eley-Rideal reaction (Section~\ref{ER}), and the unimolecular Langmuir-Hinshelwood reaction (Section~\ref{LH}). All values for the rate constants are given in the Electronic Supplementary Material.

\subsection{ DFT benchmark}\label{RDDFT}

Table~\ref{benchmark} gives an overview of the activation and reaction energies without zero-point energy {(ZPE)} corrections for reactions~\ref{R1} and~\ref{R2}. All DFT calculations comprise full geometry optimizations and all stationary points were verified by their appropriate number of imaginary frequencies, {\emph{i.e.}, zero for the reactants/products and one for the transition states. The coupled cluster values are single-point energies on MPW1B95/MG3S and M05-2X/MG3S levels for reactions~\ref{R1} and~\ref{R2}. Finally, reaction energies have also been calculated with the use of atomization energies computed with the HEAT-456QP protocol (as tabulated in \citet{Harding:2008}).

{CCSD(T) results are commonly used in computational chemistry as a gold standard for activation and reaction energies. This is, however, valid only for species where a single reference wavefuction is a good approximation. We found values of the T1 and D1 diagnostics of 0.022 and 0.062 (\ref{R1}) and 0.031 and 0.109 (\ref{R2}) in our CCSD(T)-F12 calculations. 
Thefore, additional tests with the icMRCCSD(T) method seemed important. 
{The icMRCCSD(T) method translates the accuracy of CCSD(T) to multireference cases and has been successfully applied to predicting barriers of reactions.\citep{Aoto:2016}}
An estimate of the multireference effects can be obtained from the results of icMRCCSD(T) and UCCSD(T) calculations using the cc-pVTZ basis. The computations indicate, that multireference effects only slightly lower the activation energy ($-2.8$ kJ/mol for \ref{R1} and $-1.3$ kJ/mol for \ref{R2}). We hence conclude that the multireference character of the transition state is not pronounced. This can also be seen from the contributions of the main configurations to the CASSCF wavefunction for the transition states. These are (a) a doubly occupied bonding sigma orbital with a singly occupied orbital and (b) a doubly occupied anti-bonding sigma orbital with again the singly occupied orbital. For reaction~\ref{R1} the probabilities are 0.93 and 0.03, for reaction~\ref{R2} they are 0.94 and 0.02 for configurations (a) and (b) respectively. In both cases the bonding and anti-bonding sigma orbitals are similar, albeit with different geometries for the two reactions. The orbital corresponding to the unpaired electron has contributions mainly from 1s orbital of the incoming hydrogen atom and the 2p$_\text{z}$ orbital of one of the oxygen atoms. The corresponding figures of the three orbitals for both reactions are given in the Electronic Supplementary Material. }

{Comparing the coupled cluster values to those obtained with DFT,} it is clear that the barrier of reaction~\ref{R1} is best described by the combination MPW1B95/MG3S and reaction~\ref{R2} by M05-2X/MG3S, which is in full agreement with previous findings.\citep{Ellingson:2007} 
{The slight overestimation of the barrier can result in an underestimation of the calculated reaction rate constants, but we will show that the spread in the activation energies resulting from the interaction with spectator \ce{H2O} molecules is much larger. {The need for using two different functionals for the O--O bond breaking and H-abstraction reactions is somewhat disatisfactory from a purist's point of view. At present, however, our choice is dictated by the absence of any practial functional that is good as describing both reaction paths.\citep{Ellingson:2007} The issue is alleviated to some degree by the fact that the two processes take place on different parts of the potential energy surface. } }

\begin{figure}[t]
\centering
\includegraphics[width=0.5\textwidth]{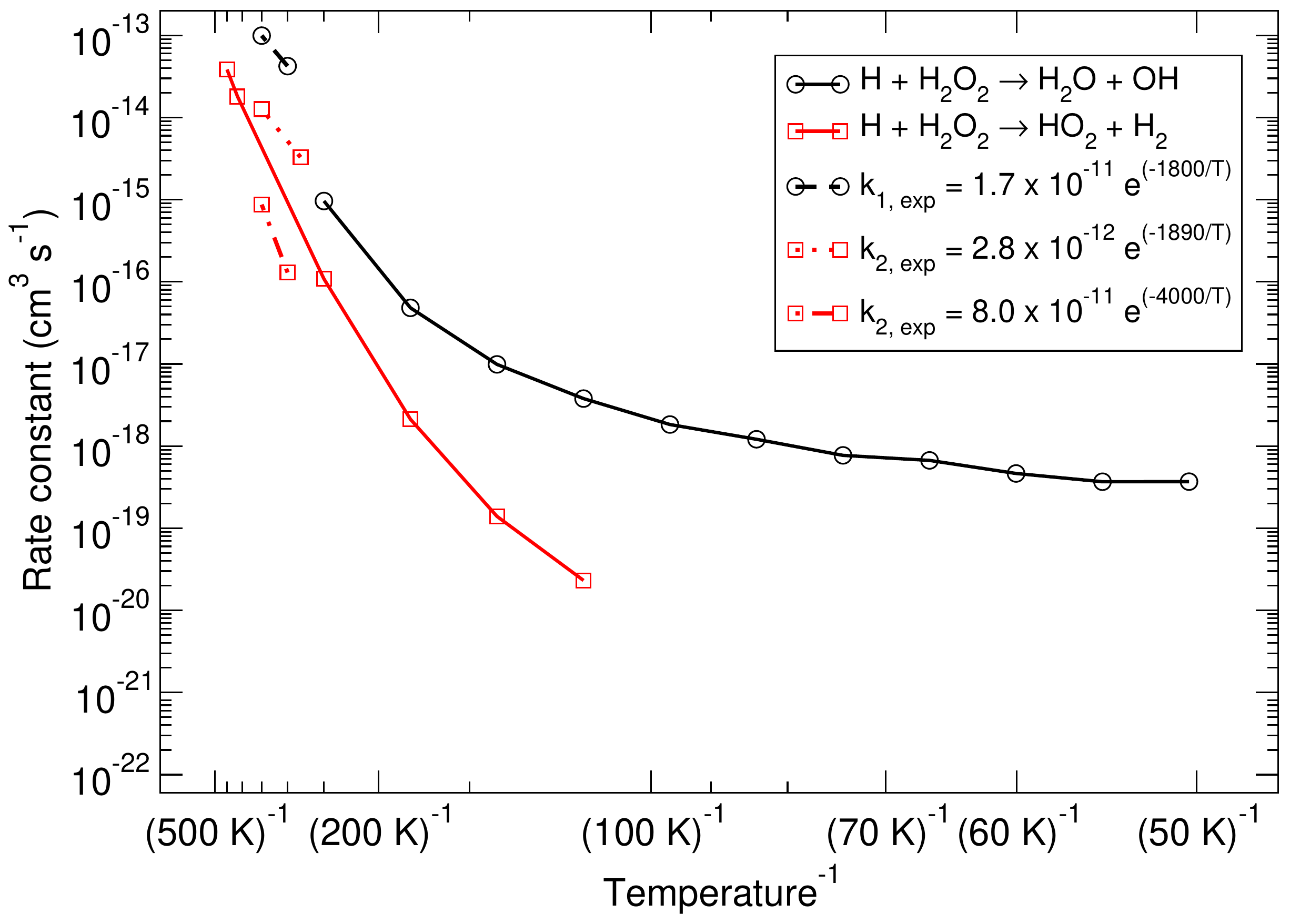}
\caption{Gas-phase branching ratio: Calculated bimolecular reaction rate constants $k_1$ and $k_2$ for reactions~\ref{R1} and~\ref{R2}, respectively, in the gas phase compared to recommended expressions derived from gas-phase experiments.\citep{Albers:1971, Klemm:1975, Baldwin:1979, Lee:1998} }
\label{GasBranching}
\end{figure}

As we will outline in Section~\ref{Gas} the main focus of the paper is on reaction~\ref{R1} and therefore we focus our attention concerning the description of dispersive interactions on this reaction only. The functional MPW1B95 has been shown to have a good performance for hydrogen bonding and weak interaction calculations.\citep{Zhao:2004} Therefore it is not obvious if an empirical dispersion correction should be applied additionally. We have tested the performance of the MPW1B95/MG3S combination with and without a D3 correction\citep{Grimme} for the \ce{H2O} dimer, trimer, and {tetramer} as well as for the interaction energy between \ce{H2O2} and \ce{H2O}. We find that without additional correction the interaction energies differ from the CCSD(T)-F12 single points by $\pm 1.6-4.3\%$, whereas with D3 correction they differ by $\pm 5.3-7.4\%$. Since a good description of the \ce{H2O} surface and the binding between \ce{H2O2} and such a surface is important for our astrochemical application, we will not use an additional dispersion correction.

\subsection{Gas-phase reaction}\label{Gas}

Figure~\ref{GasBranching} shows the bimolecular reaction rate constants calculated for reactions~\ref{R1} and~\ref{R2} for the pure gas-phase situation. Note that the black curve has been calculated with the MPW1B95/MG3S combination and the red curve with M05-2X/MG3S following earlier statements.

Instanton theory is not applicable above the crossover temperature, which are 275~ and 463~K for~\ref{R1} and~\ref{R2} respectively. This is why the two curves are cut off at the higher temperature range. From the figure the rate constants calculated at the highest temperatures can be compared to the expressions summarized by \citet{Baulch:2005} based on experimental and modeling work.\citep{Albers:1971, Klemm:1975, Baldwin:1979, Lee:1998} These expressions are recommended down to 280 or 300~K and a reasonably good correspondence to our calculated values can be seen.

\begin{figure}[t]
\centering
\includegraphics[width=0.5\textwidth]{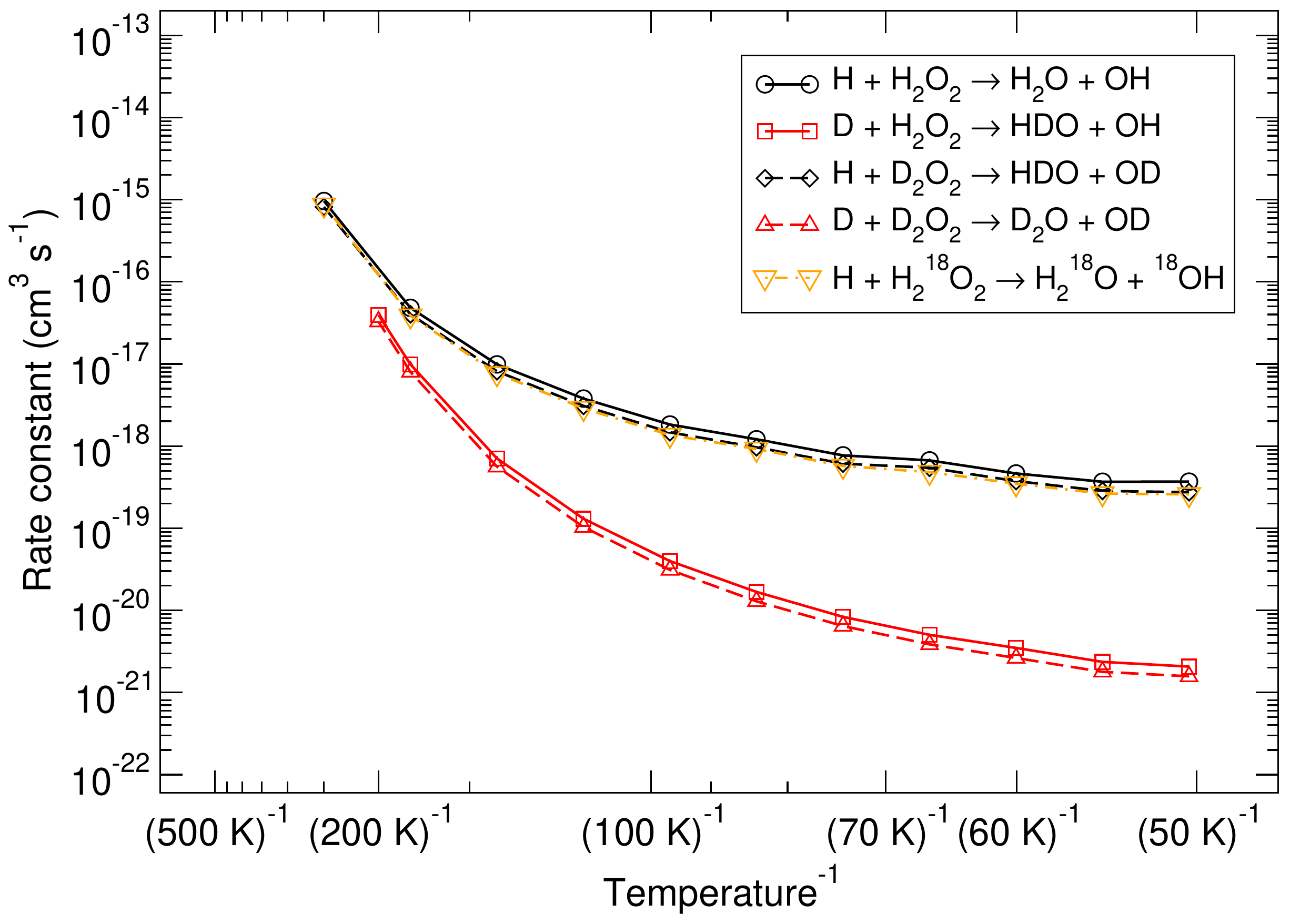}
\caption{Gas-phase kinetic isotope effect: Bimolecular reaction rate constants for reaction~\ref{R1} and isotope substituted analogues. Note that the curves for the reactions \ce{H + D2O2 -> HDO + OD} and \ce{H + H2^{18}O2 -> H2^{18}O + ^{18}OH} overlap. }
\label{GasKIE}
\end{figure}

Already at a temperature of 114~K the difference between the rate constants of~\ref{R1} and~\ref{R2} is more than two orders of magnitude. This can be explained by the much higher activation energy, despite the fact that tunneling might be expected to dominate H-abstraction at low temperatures more than the breaking of the O--O bond. Therefore, we exclude it from further investigation, \emph{i.e.}, the branching ratio \ref{R1}:\ref{R2} at lower temperature is at least 100:1. Figure~\ref{Paths} shows the instanton paths at 114~K for both reactions. The path essentially shows the delocalization of the atoms involved in the reaction. This deviates from the classical picture of overcoming a barrier and visualizes tunneling through a barrier. The lower the temperature, the more delocalization is usually visible, since tunneling then plays a larger role.

\begin{figure}[t]
\centering
\resizebox{0.23\textwidth}{!}{\includegraphics{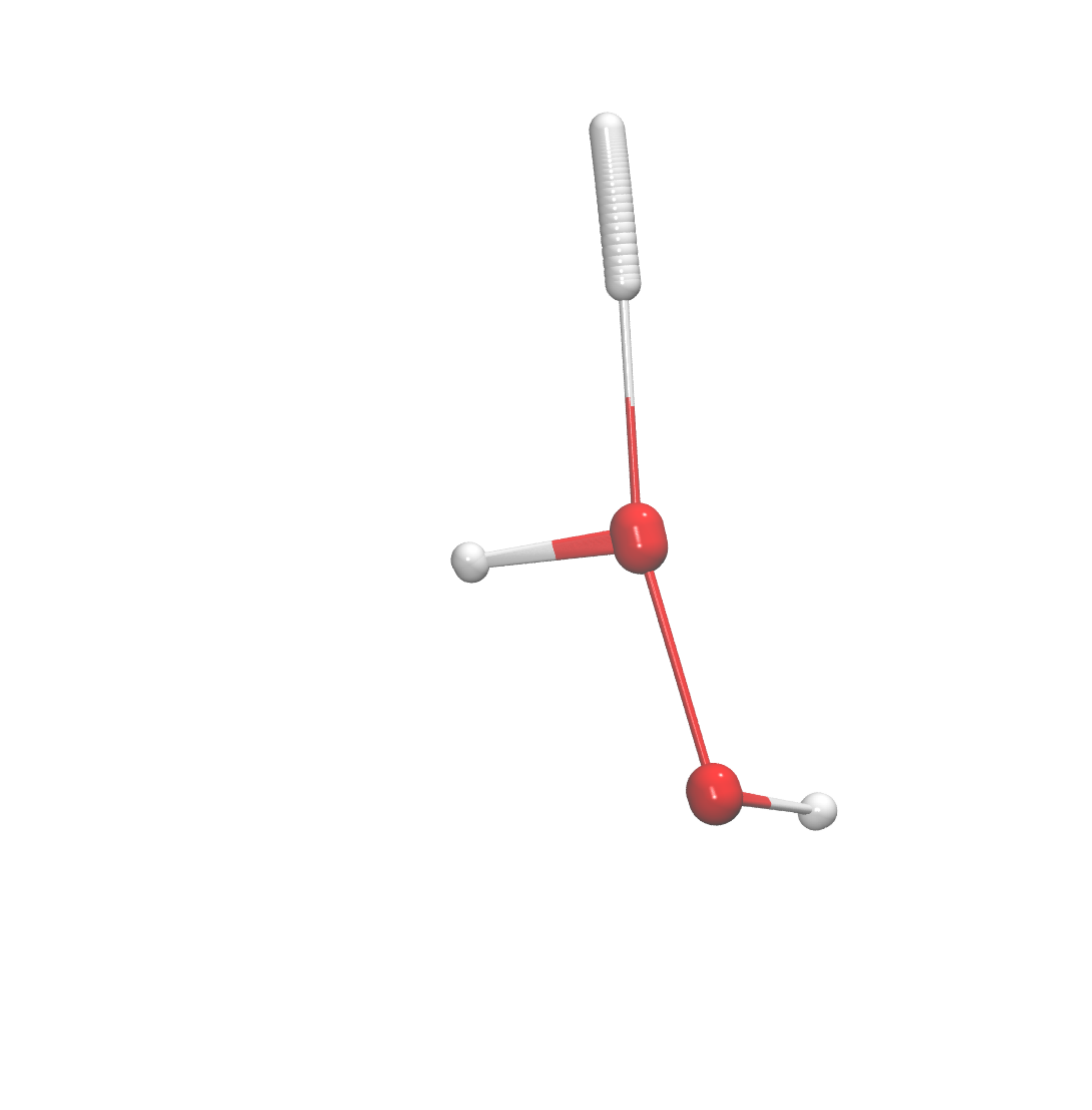}}
\resizebox{0.23\textwidth}{!}{\includegraphics{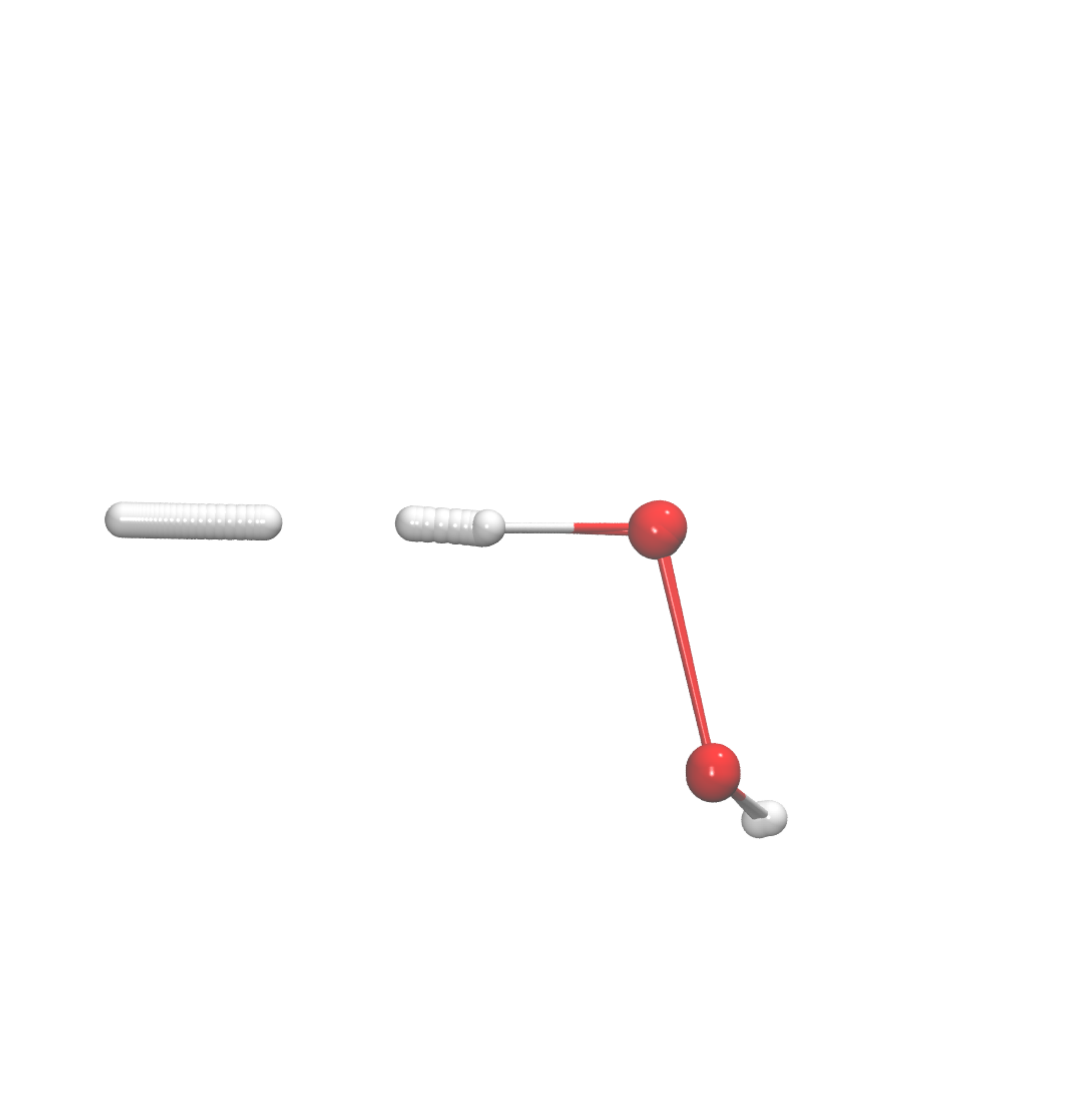}}
\caption{Instanton path for reaction~\ref{R1} (O--O bond breaking, left) and~\ref{R2} (H-abstraction, right) at 114~K. }
\label{Paths}
\end{figure}

Several possible isotopic substitutions can be made: O vs. $^{18}$O and H vs. D. In Fig.~\ref{GasKIE} the resulting rate constants for such substitutions are depicted. Firstly, changing the oxygen atom to a heavier isotope results in a lowering of the rate constant by a factor 1.4 at 50~K, because the O--O bond needs to be broken. In Fig.~\ref{Paths} it is visible why: the oxygen atoms are somewhat delocalized as well, meaning that they take part in the tunneling process. Exchanging a protium for a deuterium on the peroxide molecule has a similar small effect, \emph{i.e.}, decreasing the rate constant by a factor 1.3 at 50~K. However, for the hydrogen that approaches the peroxide and is to be added to the oxygen there is a strong kinetic isotope effect of a factor 229 at 50~K, see Table~\ref{KIE}, again visualized by strong delocalization.

\begin{figure}[t]
\centering
\includegraphics[width=0.35\textwidth]{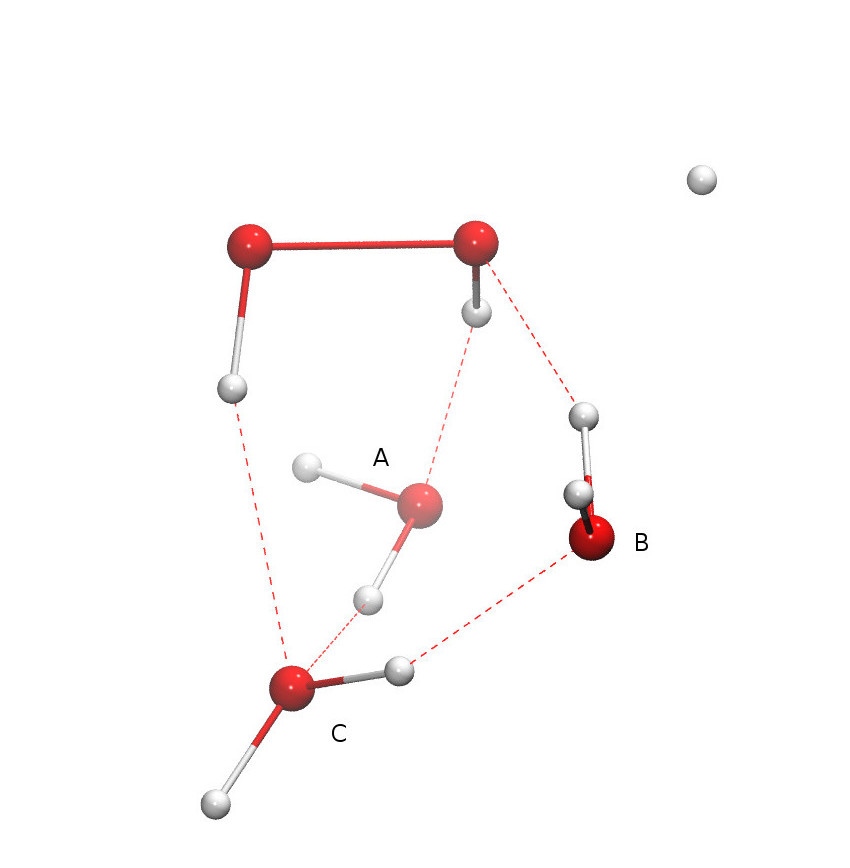}
\caption{ Transition state structure with three added water molecules and their respective labeling. }
\label{spectator}
\end{figure}

\subsection{Eley-Rideal surface reaction mechanism}\label{ER}
The Eley-Rideal reaction mechanism in surface chemistry corresponds to a reaction between one reactant that is adsorbed on a surface, and a second reactant that approaches directly from the gas-phase, hence to a bimolecular reaction between the adsorbate-surface system and the incoming atom. In our case, the adsorbate is the \ce{H2O2} molecule, the incoming atom is the H atom, and the surface is represented by spectator \ce{H2O} molecules. {The opposite case where H is pre-adsorbed and peroxide comes in from the gas phase is much less likely given the low \ce{H2O2} gas-phase abundance in the ISM.}

\begin{table}
 \centering
 \caption{ Activation energies, $E_\text{a}$ for the hydrogenation and deuteration versions of reaction~\ref{R1} with respect to two separated reactants (bimolecular ER) and to an encounter complex (unimolecular LH) in kJ/mol including zero-point energy. The cross-over temperature indicating when tunneling {dominates the reaction rate} in is given in K. }\label{ActEn}
 \begin{tabular}{llll}

 \hline
		& $E_\text{a}$ bimol.~ 	& $E_\text{a}$ unimol.~	& $T_\text{C}$ 	\\
\hline
 & \multicolumn{3}{c}{\ce{H + H2O2 -> H2O +OH}} \\
 \hline
Gas 			& 27.6 	& 26.6 	& 275 	\\
1 \ce{H2O}:\; A		& 25.1	& 24.2	& 264 	\\
1 \ce{H2O}:\; B		& 29.9	& 	& 283 	\\
2 \ce{H2O}:\; A, B	& 27.9	& 26.8  & 274 	\\
3 \ce{H2O}:\; A, B, C	& 32.6 	& 30.4	& 288 	\\
\hline
& \multicolumn{3}{c}{\ce{D + H2O2 -> HDO +OH}} \\
\hline
Gas 			& 26.2	& 25.6 	& 222 \\
2 \ce{H2O}:\; A, B	& 26.4	& 25.7	& 221 \\
3 \ce{H2O}:\; A, B, C	& 31.0	& 29.6	& 234 \\
\hline
\end{tabular}
\end{table}

\begin{table}
 \centering
 \caption{Temperature dependence of the kinetic isotope effect (Eqn.~\ref{eqnKIE}) of the LH and ER rate constants for reaction~\ref{R1} surrounded by various spectator \ce{H2O} molecules.}\label{KIE}
 \begin{tabular}{ccccccc}
 \hline
 T (K) 	& \multicolumn{3}{c}{ER bimol.} & \multicolumn{3}{c}{LH unimol.} \\
	& Gas 	& 2 \ce{H2O}	& 3 \ce{H2O}	& Gas 	& 2 \ce{H2O}	& 3 \ce{H2O}	\\
 \hline
 179	& 6.0	& 4.9		& 7.2		& 3.8	& 4.7		& 8.2 		\\
 139	& 17	& 13		& 23		& 11	& 14		& 27		\\
 114	& 36	& 26		& 50		& 24	& 30		& 64		\\
 97	& 58	& 53		& 87		& 39	& 56		& 120		\\
 84	& 91	& 87		& 146		& 63	& 84		& 215		\\
 74	& 117	& 128		& 204		& 85	& 130		& 326		\\
 66	& 169	& 148		& 266		& 127	& 176		& 461		\\
 59	& 169	& 176		& 269		& 133	& --		& 510		\\
 54	& 199	& 202		& 306		& 163	& 340		& --		\\
 50	& 229	& 202		& 522		& 197	& --		& 1194		\\
  \hline
 \end{tabular}
\end{table} 

Adding one (A or B), two (A and B), or three (A, B, and C) spectator water molecules, as depicted in Figure~\ref{spectator}, helps to elucidate the effect of the interaction between the water surface and and the reactants. In Fig.~\ref{spectator}, a transition structure with three added water molecules is shown, including their hydrogen bonded structure and respective labeling. Table~\ref{ActEn} summarizes the activation energies for the reaction between hydrogen peroxide and protium or deuterium. Adding water molecules to the reactive site indeed has an influence on the activation energies, which consequently  span a range between 25.1~and 32.5~kJ/mol. The activation energy seems to correlate with the O--O and O--H (incoming) distances in the transition state: the higher the activation energy, the larger the O--O and the smaller the O--H bond lengths, \emph{i.e.}, the later the transition state is.

The corresponding reaction rate constants are presented in Fig.~\ref{ERall}. Here, rate constants are calculated keeping the rotational partition function constant and without a symmetry factor. Therefore, the black curves without water have been recalculated and differ from those in Figs.~\ref{GasBranching} and~\ref{GasKIE}. Comparing the five curves to their respective activation energies we can see that the higher the barrier, the lower the rate constant, as can be expected. The spread at the lowest temperature, 50~K, is about 1.5 orders of magnitude. However, all curves do seem to follow the same trend, which indicates that the way the reaction proceeds it not altered by the presence of water molecules. This means that although hydrogen bonds can and are formed, the influence thereof lies only in the height of the activation energy. The rate constants seem to level off around a value of $10^{-19}$ cm$^3$s$^{-1}$ at 50~K.

\begin{figure}[t]
\centering
\includegraphics[width=0.5\textwidth]{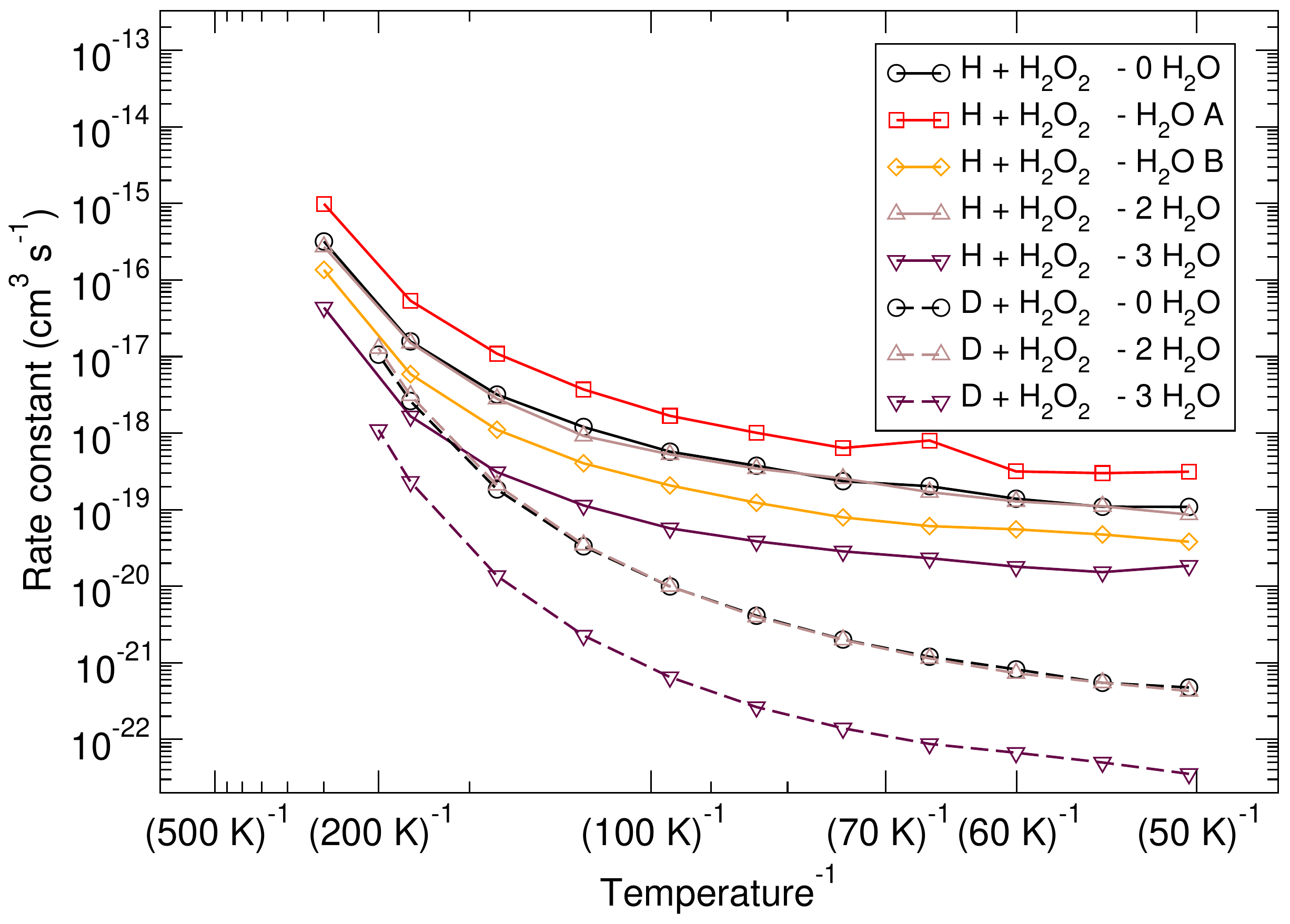}
\caption{Eley-Rideal mechanism: Bimolecular reaction rate constants for reaction~\ref{R1} surrounded by various spectator \ce{H2O} molecules and isotope substituted analogues. Note that the curves for the reactions \ce{D + H2O2} - \ce{0 H2O} and \ce{D + H2O2} - \ce{2 H2O} overlap.}
\label{ERall}
\end{figure}

Again, the same correspondence between the activation energy and rate constant is found for the deuterated reactions. Furthermore, a similar spread in the rate constants is observed at 50~K. {The ratios between the reaction rate constants for hydrogenation and deuteration of \ce{H2O2} for the ER and LH rate constants are summarized in Table~\ref{KIE}. This ratio is defined here as the kinetic isotope effect, $\phi_{\rm KIE}$, }
\begin{equation}
\phi_{\rm KIE} \equiv \frac{k_{\ce{H + H2O2}}}{k_{\ce{D + H2O2}}} \;. \label{eqnKIE}
\end{equation}

\subsection{Langmuir-Hinshelwood surface reaction mechanism}\label{LH}
The Langmuir-Hinshelwood reaction mechanism assumes two already adsorbed species on the surface that find each other via diffusion, sit next to each other, \emph{i.e.}, form a pre-reactive complex, and then attempt a reaction. This is thus related to a unimolecular reaction starting out from the encounter complex; note the different units.

Table~\ref{ActEn} gives the activation energies for the reactions {with respect to} this encounter complex. It is clear from the small differences in the barriers between the first and the second column of the table that the interaction between the hydrogen atom and the \ce{H2O2}(+(\ce{H2O})$_n$) system is very weak. In fact the interaction energies are slightly positive, 1--2 kJ/mol. This {could be} caused by the poor description of the Van der Waals interaction by DFT, however, {even} an additional D3 correction {leaves} the interaction energies positive. The result is that at lower temperatures the rate constants become more noisy. This weak interaction of the H atoms shows that it is not trivial to define an encounter complex, since most likely one should take an ensemble of possible configurations into account. This will be dealt with in a future study.

\begin{figure}[t]
\centering
\includegraphics[width=0.49\textwidth]{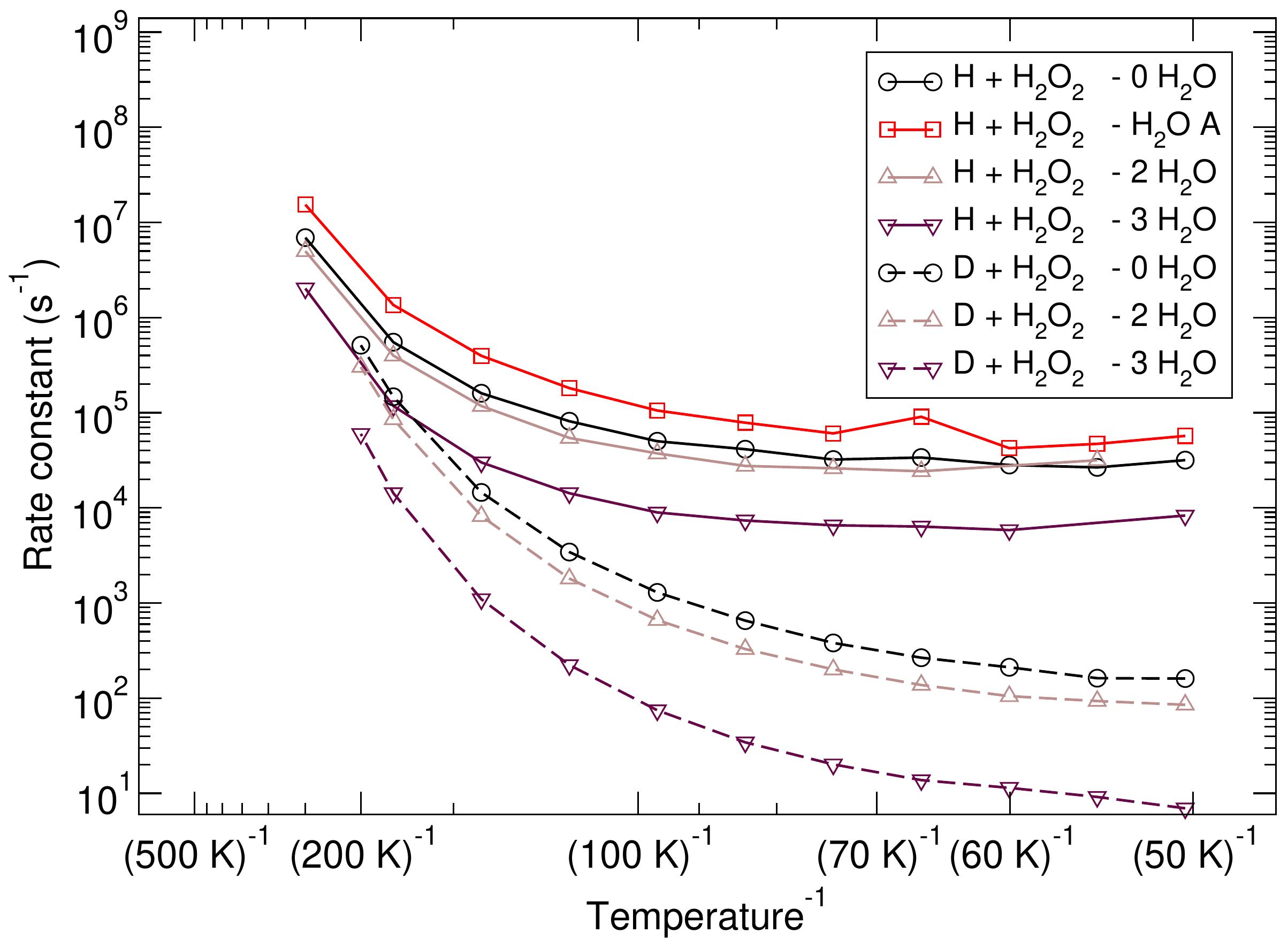}
\caption{Langmuir-Hinshelwood mechanism: Unimolecular reaction rate constants for reaction~\ref{R1} surrounded by various spectator \ce{H2O} molecules and isotope substituted analogues. }
\label{LHall}
\end{figure}

The reaction rate constants are presented in Fig.~\ref{LHall}. Again the rotational partition function has been kept constant. Furthermore, a similar trend as for the ER bimolecular case can be observed, with a spread of about an order of magnitude. The rate constants seem to level off around a value of $2\times10^{4}$ s$^{-1}$ at 50~K.

Experimentally, under ultra-high vacuum conditions, it is assumed that the LH mechanism plays the dominant role. Early work by \citet{Miyauchi:2008, Ioppolo:2008, Cuppen:2010B} already specifically mentions the role of the reaction \ce{H + H2O2 -> H2O + OH} within the water network, but it was only by \citet{Oba:2014} that the reaction and the role of tunneling were explored to a greater detail. In their paper, they show a clear kinetic isotope effect between hydrogenation and deuteration of \ce{H2O2} and also of \ce{D2O2}. Since experimentally it is not possible to determine the amount of H/D atoms residing on the surface, the reported $\phi_{\rm KIE}$'s are based on effective rate constants and have a value of approximately 50. A more exact {treatment} would have to include the exact H:D flux ratio as well as diffusion and recombination rates. Our low-temperature $\phi_{\rm KIE}$ is at least a factor of four higher, which is caused by the fact that the theoretical $\phi_{\rm KIE}$ is based solely on the rate constant for the reactions, excluding any additional effects. {Since the longer lifetime of D atoms on the surface with respect to that of H atoms on the surface indeed is expected to lower the kinetic isotope effect}, the agreement to experiments is actually reasonable. Note that in the unimolecular case, the addition of spectator molecules strongly influences {the kinetic isotope effect, the reason for this is at present unclear}.

\begin{figure}[t]
\centering
\includegraphics[width=0.5\textwidth]{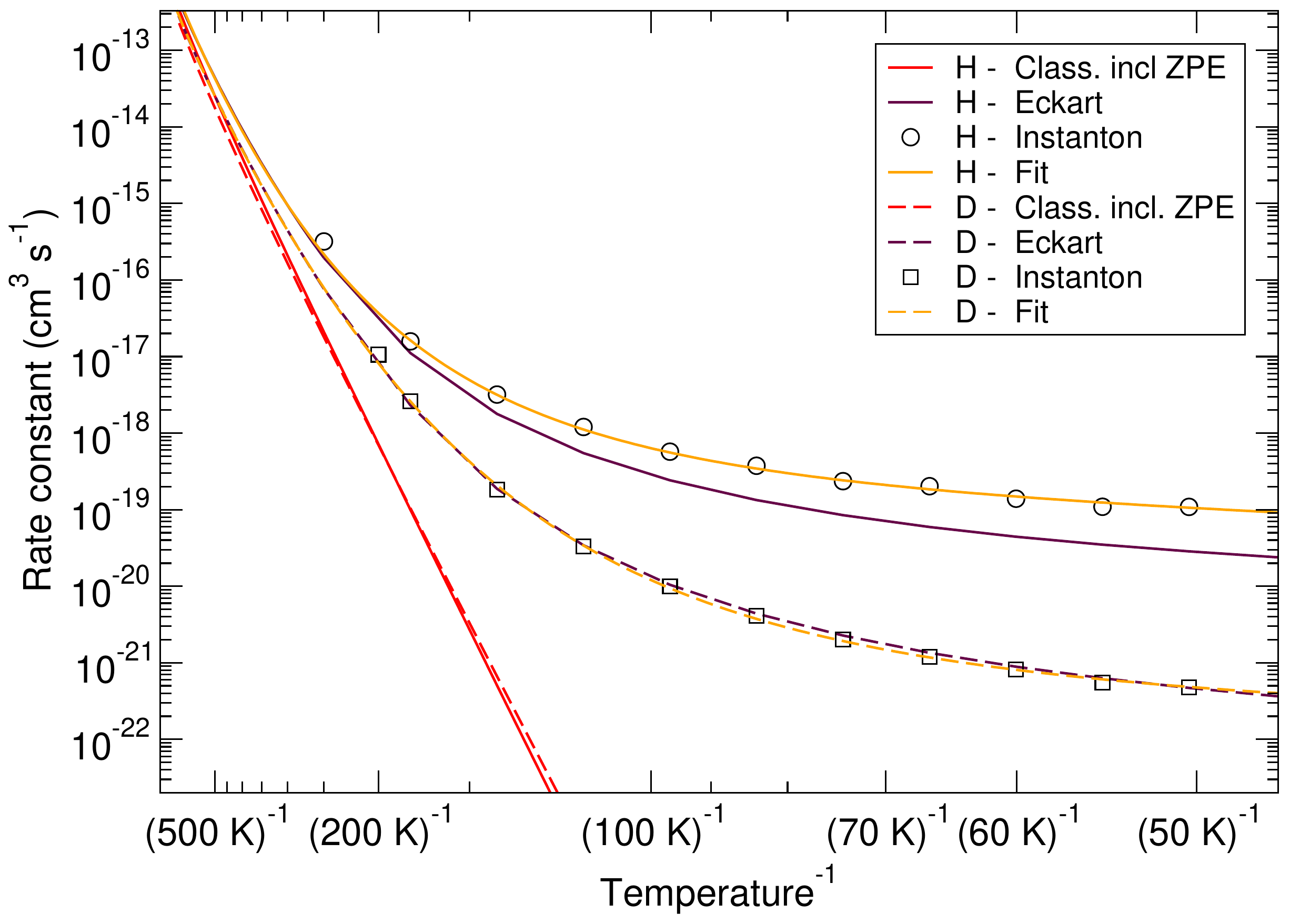}
\caption{Eley-Rideal mechanism: Comparison between rate constants calculated with transition state theory including quantized vibrations, the Eckart approximation, and instanton theory. The fit to Eqn~\ref{Zheng} is also shown. }
\label{ERclass}
\end{figure}

\section{Implementation in astrochemical models}\label{AstroIm}
The rate constants that are presented in this work can be used as input for astrochemical models. To that purpose, we have fitted rate constants to equation~\ref{Zheng} with fitting parameters $\alpha$, $\beta$, $\gamma$, and $T_0$. The values of these parameters are given in Table~\ref{fits}. Note that in all cases, we used the pure gas-phase geometries, \emph{i.e.} without spectator molecules, but calculated the ER and LH rate constants keeping the rotation partition function constant. In Fig.~\ref{ERall} we showed that the rate constants without spectator molecules lie nicely within the range spanned by the various geometries and can thus be interpreted as an average value. The full spread is about 1--1.5 orders of magnitude and therefore the exact values of the rate constant fits should be considered to have an uncertainty of approximately a factor 5. We stress here that in astrochemical grain-surface reaction modeling this accuracy is in fact quite sufficient. Uncertainties of that order can even be found in gas-phase reaction networks although these are typically better constained.\citep{wakelam:2005}

As outlined above, this reaction can, in principal, take place both in the gas phase and on a (water) surface. As a result of the low interstellar abundances of \ce{H2O2} in the gas phase, the gas-phase route is unlikely.
A gas-phase hydrogen atom could directly strike a reaction partner that is adsorbed to a water or grain surface, as in a bimolecular process. However, the timescales involved in dark cloud chemistry are large enough for a hydrogen atom to scan the surface and meet a \ce{H2O2} molecule. Therefore, unimolecular rate constants have a direct correspondence to the process as it occurs in the interstellar medium. For the sake of completeness we provide fit parameters for all three possibilities. 

The rate constants presented here are not directly comparable to the expressions commonly used in astrochemical modelling. Usually in such models the tunneling probability, $P_\text{tunn.}$, is described by the rectangular barrier approximation
\begin{equation}
 P_\text{tunn.} = \exp\left(\frac{-2a}{\hbar}\sqrt{2\mu E_\text{a}}\right) 
\end{equation}
where $a$ is interpreted as the barrier width, $\mu$ as the effective mass and $E_\text{a}$ the activation energy of the reaction including ZPE. This is a very convenient expression, because is it implementable in rate equation models. Most of these take diffusion into account in the calculation of the LH reaction rate:
\begin{equation}
 R_\text{LH, react.} = P_\text{react.} \; R_\text{diffusion} \;.
\end{equation}
Here, the rate of diffusion, {$R_\text{diffusion}$}, includes the diffusion of both species, 
\begin{equation}
R_\text{diffusion} = \frac{k_\text{diff, A}+k_\text{diff, B}}{N_\text{sites}}n_\text{A}n_\text{B}                                                                 
\end{equation}
with $k_\text{diff}$ the unimolecular diffusion rate constant, $N_\text{sites}$ the number of surface sites and $n_\text{X}$ the concentration of species X. The probability for reaction is composed of the competition between reaction, diffusion out of the site, and desorption {rate constants}, 
\begin{equation}
P_\text{react.} = \frac{k_\text{react.}}{k_\text{react.} + k_\text{diff.} + k_\text{desorp.}} \;.
\end{equation}
In models, at low temperature, $k_\text{react.}$ is often approximated by a trial frequency, $\nu$, multiplied by $P_\text{tunn.}$, whereas what we calculated corresponds directly to $k_\text{react.} = k_\text{tunn.}$.  Therefore, the best way to compare the instanton rate constants to the rectangular barrier approximation is to look at the kinetic isotope effect, see Table~\ref{CompKIE}. Furthermore, \citet{Taquet:2013} improved on the rectangular barrier approximation by the use of an Eckart correction to describe tunneling. In the same Table~\ref{CompKIE} the $\phi_{\rm KIE}$'s calculated with a symmetric Eckart approximation are given as well. It is clear that both the rectangular barrier and the Eckart approximation are inadequate descriptions of tunneling at low temperatures.

\begin{figure}[t]
\centering
\includegraphics[width=0.49\textwidth]{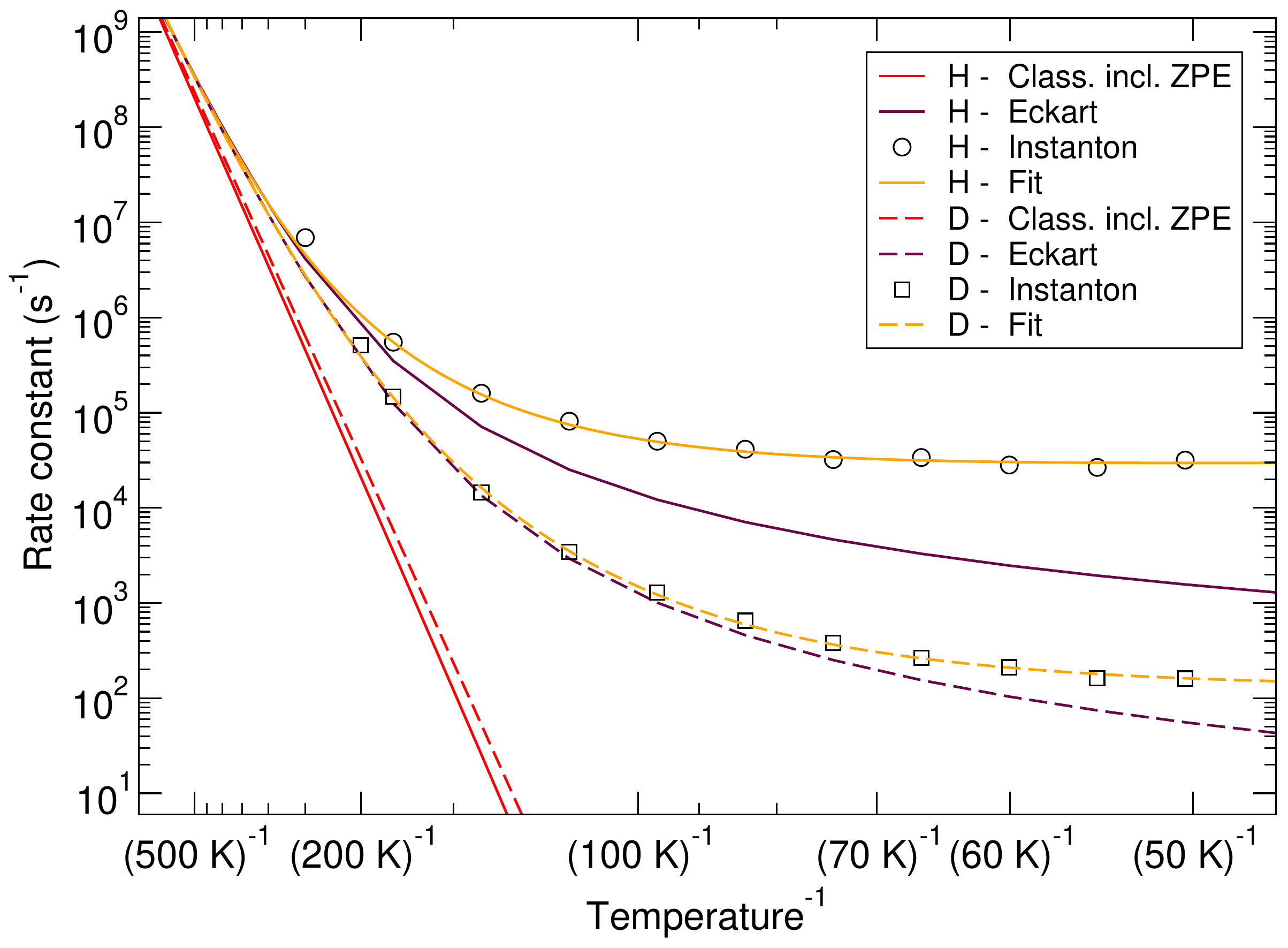}
\caption{Langmuir-Hinshelwood mechanism: Comparison between rate constants calculated with transition state theory including quantized vibrations, the Eckart approximation, and instanton theory. The fit to Eqn~\ref{Zheng} is also shown. }
\label{LHclass}
\end{figure}

Figs.~\ref{ERclass} and~\ref{LHclass} show a comparison of the rate constants as calculated with harmonic transition state theory (incl. ZPE), with an Eckart approximation, and with the instanton method. These figures again indicate that any correspondence between the Eckart approximation and instanton rates is only fortuitous, compare the ER bimolecular \ce{D + H2O2} rate constants {(which overlaps with the instanton curve)} with those for the LH unimolecular \ce{H + H2O2} case. More generally speaking, Eckart rate constants are likely to underestimate the true value and show a wrong low-temperature behavior.

\begin{table}
 \centering
 \caption{Parameters fitted down to 50~K to describe the reactions \ce{H + H2O2} and \ce{D + H2O2} in the gas phase, with an ER, and an LH mechanism. }\label{fits}
 \begin{tabular}{llll}
 \hline
 Parameter & Gas  & ER  & LH  \\
\hline
 & \multicolumn{3}{c}{\ce{H + H2O2 -> H2O + OH}} \\
 \hline
 $\alpha$ (cm$^{3}$ s$^{-1}$ / s$^{-1}$) 	&   1.92 $\times$ $10^{-12}$	& 2.74 $\times$ $10^{-13}$ 	& 1.51  $\times$ $10^{10}$ \\
 $\beta$ 		& 2.54 & 2.61 & 0.86 \\
 $\gamma$ (K) 		& 1660 & 1630 & 1750 \\
 $T_0$ (K) 		& 180 & 180 & 180 \\
 \hline
  & \multicolumn{3}{c}{\ce{D + H2O2 -> HDO + OH}} \\
\hline
 $\alpha$ (cm$^{3}$ s$^{-1}$ / s$^{-1}$) 	&  1.09 $\times$ $10^{-12}$ & 2.76  $\times$ $10^{-13}$ 	& 8.37  $\times$ $10^{9}$ \\
 $\beta$ 		& 2.65 & 2.69 & 1.19 \\
 $\gamma$ (K) 		& 1615 & 1600 & 1625 \\
 $T_0$ (K) 		& 125 & 125 & 125 \\
 \hline
 \end{tabular}
\end{table}

\begin{table}
 \centering
 \caption{Kinetic isotope effect: Comparison of the $\phi_{\rm KIE}$ calculated with instanton theory, the Eckart approximation, and the rectangular barrier approximation at 50~K.}\label{CompKIE}
 \begin{tabular}{llll}
  \hline
  & Instanton & Eckart & Rect. barrier \\
  \hline
ER & 197 & 27 & 6945 \\
LH & 229 & 60 & 7033 \\
  \hline
 \end{tabular}
\end{table}

\section{Conclusion}\label{Concl}
With the study of the reaction between hydrogen and hydrogen peroxide the final step of the water formation sequence starting from molecular oxygen is further quantified. In particular, attention is paid to the low-temperature behavior, the kinetic isotope effect, and the influence of spectator water molecules mimicking an icy grain surface.

Specifically, 
\begin{itemize}
 \item prior to calculating rate constants the method of choice (DFT) has been benchmarked to single-reference and multireference coupled cluster single-point energies for the stationary points on the potential energy surface,
 \item a branching ratio between the rate constants for O--O bond breaking, \ref{R1}, and H-abstraction, \ref{R2}, of at least 100:1 was established at a temperature of 114~K,
 \item rate constants that apply to the gas-phase, surface Eley-Rideal, and surface Langmuir-Hinshelwood mechanisms are now available down to 50~K,
 \item the 50~K results, or an extrapolation down to at least 30~K via the fitted expression Eqn~\ref{Zheng} (Table~\ref{fits}), can be used as a reasonable guess for even lower-temperature surface processes thanks to the rate constants leveling off with decreasing temperature,
 \item {quantitative agreement with experimental gas-phase data and qualitative agreement with the experimental surface kinetic isotope effects is found,}
 \item the addition of spectator molecules indeed influences on the reaction rate constant and kinetic isotope effect, mainly by influencing the transition state structure, which leads to a change in the activation energy of the reaction,
 \item general trends {such as the asymptotic behavior of the curves}, don't seem to be strongly affected {by the addition of spectator molecules}, but it is important to note that the surface aids in bringing the two reactants together and allows for heat dissipation of the exothermicity,
 \item a comparison between the rectangular barrier approximation, the Eckart approximation, and instanton rate constants shows that both approximations leads to large errors (more than an order of magnitude).
\end{itemize}
The quantification of the reaction rate constants for \ce{H (D) + H2O2 -> H2O (HDO) + OH} can help to constrain the \ce{HDO}/\ce{H2O} ratio in the ice. Note that the high $\phi_{\rm KIE}$'s of more than 200 (Table~\ref{KIE}) indicate that a high abundance of \ce{H2O2} results in a decrease of the \ce{HDO}/\ce{H2O}. Furthermore it is of aid to elucidate the \ce{H2O2} abundance on surfaces.

\section{Acknowledgements}
TL and JK are financially supported by the European Union's Horizon 2020 research and innovation programme (grant agreement No. 646717, TUNNELCHEM). PKS and AK would like to thank the German Research Foundation (DFG) for financial support of the project within the Cluster of Excellence in Simulation Technology (EXC 310/2) at the University of Stuttgart. Jan Meisner and Sonia \'Alvarez Barcia are acknowledged for their help and discussions in various stages of the project. 



\singlespacing
\bibliography{thanja.bib} 
\bibliographystyle{rsc} 

\end{document}